\documentclass[12pt]{article}
\usepackage{graphics,graphicx}
\usepackage{amssymb,epsfig,amsmath,euscript,array}
\usepackage{cite}

\makeatletter
\@addtoreset{equation}{section}
\makeatother

\newcounter{multieqs}

\newcommand{\be}{\begin{equation}}
\newcommand{\ee}{\end{equation}}

\newcommand{\bm}[1]{\mbox{\boldmath $#1$}}

\newcommand{\kslash}{k \!\!\! / }

\newcommand{\lslash}{l \!\! / }
\newcommand{\Pslash}{P \!\!\!\! / }

\newcommand{\islash}{i \!\!\! / }
\newcommand{\jslash}{j \!\!\! / }
\newcommand{\aslash}{a \!\!\! / }
\newcommand{\bslash}{{b \hspace{-6pt} \slash} }

\newcommand{\onslash}{1 \!\!\! / }
\newcommand{\twslash}{2 \!\!\!/ }
\newcommand{\thslash}{3 \!\!\!/ }
\newcommand{\foslash}{4 \!\!\! / }
\newcommand{\fislash}{5 \!\!\! / }

\newcommand{\mslash}{m \!\!\! / }

\def\bd{\begin{document}}
\def\ed{\end{document}}
\def\nn{\nonumber}
\def\bea{\begin{eqnarray}}
\def\eea{\end{eqnarray}}

\def\ab{(ijab)}
\def\ba{(ijba)}
\def\ijab{{\tr}_{+}(\islash\, \jslash\, \aslash \, \bslash)}
\def\ijba{{\tr}_{+}(\islash\, \jslash\, \bslash \, \aslash)}
\def\ijaP{{\tr}_{+}(\islash\, \jslash\, \aslash \, \Pslash)}
\def\ijPLa{{\tr}_{+}(\islash\, \jslash\, \Pslash_L \, \aslash)}
\def\ijaPL{{\tr}_{+}(\islash\, \jslash\, \aslash \, \Pslash_L)}
\def\ijPLza{{\tr}_{+}(\islash\, \jslash\, \Pslash_{L;z} \, \aslash)}
\def\ijaPLz{{\tr}_{+}(\islash\, \jslash\, \aslash \, \Pslash_{L;z})}
\def\ijPa{{\tr}_{+}(\islash\, \jslash\, \Pslash \, \aslash)}
\def\iaPb{{\tr}_{+}(\islash\, \aslash\, \Pslash \, \bslash)}
\def\ibPa{{\tr}_{+}(\islash\, \bslash\, \Pslash \, \aslash)}
\def\ijPmu{{\tr}_{+}(\islash\, \jslash\, \Pslash \, \mu)}
\def\ibmuP{{\tr}_{+}(\islash\, \bslash\, \mu \, \Pslash)}
\def\ibmua{{\tr}_{+}(\islash\, \bslash\, \mu \, \aslash)}
\def\iamub{{\tr}_{+}(\islash\, \aslash\, \mu \, \bslash)}
\def\jaPb{{\tr}_{+}(\jslash\, \aslash\, \Pslash \, \bslash)}
\def\ijmuP{{\tr}_{+}(\islash\, \jslash\, \mu \, \Pslash)}
\def\ijmum{{\tr}_{+}(\islash\, \jslash\, \mu \, \mslash)}
\def\ijmmu{{\tr}_{+}(\islash\, \jslash\, \mslash \, \mu)}
\def\ijmP{{\tr}_{+}(\islash\, \jslash\, \mslash \, \Pslash)}
\def\iabP{{\tr}_{+}(\islash\, \aslash\, \bslash \, \Pslash)}
\def\ijbP{{\tr}_{+}(\islash\, \jslash\, \bslash \, \Pslash)}
\def\jbPa{{\tr}_{+}(\jslash\, \bslash\, \Pslash \, \aslash)}
\def\ijPb{{\tr}_{+}(\islash\, \jslash\, \Pslash \, \bslash)}
\def\jbmua{{\tr}_{+}(\jslash\, \bslash\, \mu \, \aslash)}

\def\loablt{ {\tr}_{+}(\lslash_1\, \aslash \, \bslash\, \lslash_2)}

\def\ijlolt{{\tr}_{+}(\islash\, \jslash\, \lslash_1 \, \lslash_2)}
\def\ijltlo{{\tr}_{+}(\islash\, \jslash\, \lslash_2 \, \lslash_1)}
\def\ibloa{{\tr}_{+}(\islash\, \bslash\, \lslash_1 \, \aslash)}
\def\jaltb{{\tr}_{+}(\jslash\, \aslash\, \lslash_2 \, \bslash)}
\def\ialtb{{\tr}_{+}(\islash\, \aslash\, \lslash_2 \, \bslash)}
\def\bltloa{{\tr}_{+}(\bslash\, \lslash_2\, \lslash_1 \, \aslash)}
\def\jbloa{{\tr}_{+}(\jslash\, \bslash\, \lslash_1 \, \aslash)}
\def\ibPb{{\tr}_{+}(\islash\, \bslash\, \Pslash \, \bslash)}
\def\ijltb{{\tr}_{+}(\islash\, \jslash\, \lslash_2 \, \bslash)}

\def\ijloa{{\tr}_{+}(\islash\, \jslash\,  \lslash_1 \, \aslash)}
\def\ijblt{{\tr}_{+}(\islash\, \jslash\,  \bslash \, \lslash_2)}

\def\jakb{{\tr}_{+}(\jslash\, \aslash\, \kslash \, \bslash)}
\def\iakb{{\tr}_{+}(\islash\, \aslash\, \kslash \, \bslash)}

\def\tofo{{\tr}_{+}(\onslash\, \thslash\, \twslash \, \foslash)}
\def\foto{{\tr}_{+}(\onslash\, \thslash\, \foslash \, \twslash)}
\def\tofi{{\tr}_{+}(\onslash\, \thslash\, \twslash \, \fislash)}
\def\fito{{\tr}_{+}(\onslash\, \thslash\, \fislash \, \twslash)}

\def\lrangle#1#2{\langle #1\,#2\rangle}

\def\Li{{$\rm Li}_2$}
\def\eps{\epsilon}
\def\epsuv{{\epsilon_{\rm \mbox{\tiny UV}}}}
\let\bm=\bibitem
\let\la=\label

\def\npb#1#2#3{Nucl. Phys. {\bf{B#1}} #3 (#2)}
\def\plb#1#2#3{Phys. Lett. {\bf{#1B}} #3 (#2)}
\def\prl#1#2#3{Phys. Rev. Lett. {\bf{#1}} #3 (#2)}
\def\prd#1#2#3{Phys. Rev. {D \bf{#1}} #3 (#2)}
\def\cmp#1#2#3{Comm. Math. Phys. {\bf{#1}} #3 (#2)}
\def\cqg#1#2#3{Class. Quantum Grav. {\bf{#1}} #3 (#2)}
\def\nppsa#1#2#3{Nucl. Phys. B (Proc. Suppl.) {\bf{#1A}}#3 (#2)}
\def\ap#1#2#3{Ann. of Phys. {\bf{#1}} #3 (#2)}
\def\ijmp#1#2#3{Int. J. Mod. Phys. {\bf{A#1}} #3 (#2)}
\def\rmp#1#2#3{Rev. Mod. Phys. {\bf{#1}} #3 (#2)}
\def\mpla#1#2#3{Mod. Phys. Lett. {\bf A#1} #3 (#2)}
\def\jhep#1#2#3{J. High Energy Phys. {\bf #1} #3 (#2)}
\def\atmp#1#2#3{Adv. Theor. Math. Phys. {\bf #1} #3 (#2)}
\newcommand{\EQ}[1]{\begin{equation} #1 \end{equation}}
\newcommand{\AL}[1]{\begin{subequations}\begin{align} #1 \end{align}\end{subequations}}
\newcommand{\SP}[1]{\begin{equation}\begin{split} #1 \end{split}\end{equation}}
\newcommand{\ALAT}[2]{\begin{subequations}\begin{alignat}{#1} #2 \end{alignat}
                        \end{subequations}}
\def\beqa{\begin{eqnarray}}
\def\eeqa{\end{eqnarray}}
\def\beq{\begin{equation}}
\def\eeq{\end{equation}}
\def\sst{\scriptscriptstyle}
\def\thetabar{\bar\theta}
\def\Tr{{\rm Tr}}
\def\one{\mbox{1 \kern-.59em {\rm l}}}
 \def\Nh{\hat{N}}

\def\a{\alpha}      \def\da{{\dot\alpha}}
\def\b{\beta}       \def\db{{\dot\beta}}
\def\c{\gamma}  \def\G{\Gamma}  \def\cdt{\dot\gamma}
\def\d{\delta}  \def\D{\Delta}  \def\ddt{\dot\delta}
\def\e{\epsilon}        \def\vare{\varepsilon}
\def\f{\phi}    \def\F{\Phi}    \def\vvf{\f}
\def\h{\eta}
\def\k{\kappa}
\def\l{\lambda} \def\L{\Lambda}
\def\m{\mu} \def\n{\nu}
\def\o{\omega}
\def\p{\pi} \def\P{\Pi}
\def\r{\rho}
\def\s{\sigma}  \def\S{\Sigma}
\def\t{\tau}
\def\th{\theta} \def\Th{\Theta} \def\vth{\vartheta}
\def\X{\Xeta}
\def\z{\zeta}

\def\cA{{\cal A}} \def\cB{{\cal B}} \def\cC{{\cal C}}
\def\cD{{\cal D}} \def\cE{{\cal E}} \def\cF{{\cal F}}
\def\cG{{\cal G}} \def\cH{{\cal H}} \def\cI{{\cal I}}
\def\cJ{{\cal J}} \def\cK{{\cal K}} \def\cL{{\cal L}}
\def\cM{{\cal M}} \def\cN{{\cal N}} \def\cO{{\cal O}}
\def\cP{{\cal P}} \def\cQ{{\cal Q}} \def\cR{{\cal R}}
\def\cS{{\cal S}} \def\cT{{\cal T}} \def\cU{{\cal U}}
\def\cV{{\cal V}} \def\cW{{\cal W}} \def\cX{{\cal X}}
\def\cY{{\cal Y}} \def\cZ{{\cal Z}}

\def\ua{\underline{\alpha}}
\def\ub{\underline{\phantom{\alpha}}\!\!\!\beta}
\def\uc{\underline{\phantom{\alpha}}\!\!\!\gamma}
\def\um{\underline{\mu}}
\def\ud{\underline\delta}
\def\ue{\underline\epsilon}
\def\una{\underline a}\def\unA{\underline A}
\def\unb{\underline b}\def\unB{\underline B}
\def\unc{\underline c}\def\unC{\underline C}
\def\und{\underline d}\def\unD{\underline D}
\def\une{\underline e}\def\unE{\underline E}
\def\unf{\underline{\phantom{e}}\!\!\!\! f}\def\unF{\underline F}
\def\unm{\underline m}\def\unM{\underline M}
\def\unn{\underline n}\def\unN{\underline N}
\def\unp{\underline{\phantom{a}}\!\!\! p}\def\unP{\underline P}
\def\unq{\underline{\phantom{a}}\!\!\! q}
\def\unQ{\underline{\phantom{A}}\!\!\!\! Q}
\def\unH{\underline{H}}

\def\As {{A \hspace{-6.4pt} \slash}\;}
\def\bs {{b \hspace{-6.4pt} \slash}\;}
\def\Ds {{D \hspace{-6.4pt} \slash}\;}
\def\ds {{\del \hspace{-6.4pt} \slash}\;}
\def\ss {{\s \hspace{-6.4pt} \slash}\;}
\def\ks {{ k \hspace{-6.4pt} \slash}\;}
\def\ps {{p \hspace{-6.4pt} \slash}\;}
\def\pas {{{p_1} \hspace{-6.4pt} \slash}\;}
\def\pbs {{{p_2} \hspace{-6.4pt} \slash}\;}
\def\Ps {{P \hspace{-6.4pt} \slash}\;}
\def\Qs {{Q \hspace{-6.4pt} \slash}\;}

\def\Fh{\hat{F}}
\def\Vh{\hat{V}}
\def\Xh{\hat{X}}
\def\ah{\hat{a}}
\def\xh{\hat{x}}
\def\yh{\hat{y}}
\def\ph{\hat{p}}
\def\xih{\hat{\xi}}
\def\psit{\tilde{\psi}}
\def\Psit{\tilde{\Psi}}
\def\tht{\tilde{\th}}
\def\lt{\tilde{\lambda}}
\def\llt{\tilde{l}}
\def\At{\tilde{A}}
\def\Qt{\tilde{Q}}
\def\Rt{\tilde{R}}
\def\Nt{\tilde{N}}

\def\at{\tilde{a}}
\def\st{\tilde{s}}
\def\ft{\tilde{f}}
\def\pt{\tilde{p}}
\def\qt{\tilde{q}}
\def\vt{\tilde{v}}
\def\nt{\tilde{n}}

\def\delb{\bar{\partial}}
\def\bz{\bar{z}}
\def\bD{\bar{D}}
\def\bB{\bar{B}}

\def\bk{{\bf k}}
\def\bl{{\bf l}}
\def\bp{{\bf p}}
\def\bq{{\bf q}}
\def\br{{\bf r}}
\def\bx{{\bf x}}
\def\by{{\bf y}}
\def\bR{{\bf R}}
\def\bV{{\bf V}}

\def\d{\delta}\def\D{\Delta}\def\ddt{\dot\delta}
\def\pa{\partial} \def\del{\partial}
\def\xx{\times}
\def\uno{\mbox{1 \kern-.59em {\rm l}}}
\def\trp{^{\top}}
\def\inv{^{-1}}
\def\dag{{^{\dagger}}}
\def\pr{^{\prime}}
\def\lan{\langle}
\def\ran{\rangle}
\def\rar{\rightarrow}
\def\lar{\leftarrow}
\def\lrar{\leftrightarrow}
\newcommand{\0}{\,\!}      
\def\one{1\!\!1\,\,}
\def\im{\imath}
\def\jm{\jmath}
\newcommand{\tr}{\mbox{tr}}
\newcommand{\slsh}[1]{/ \!\!\!\! #1}
\def\vac{|0\rangle}
\def\lvac{\langle 0|}
\def\hlf{\frac{1}{2}}
\def\ove#1{\frac{1}{#1}}
\def\Box{\square}
\def\ZZ{\mathbb{Z}}
\def\CC#1{({\bf #1})}
\def\bcomment#1{}
\def\bfhat#1{{\bf \hat{#1}}}
\def\VEV#1{\left\langle #1\right\rangle}
\newcommand{\ex}[1]{{\rm e}^{#1}} \def\ii{{\rm i}}
\def\rr{{\rm r}} \def\rs{{\rm s}}\def\rv{{\rm v}}
\def\ri{{\rm i}}\def\rj{{\rm j}}
\newcommand{\lrbrk}[1]{\left(#1\right)}
\newcommand{\sfrac}[2]{{\textstyle\frac{#1}{#2}}}

\def\Li{{\rm Li}_2}

\font\mybb=msbm10 at 12pt
\def\bb#1{\hbox{\mybb#1}}

\font\myBB=msbm10 at 18pt
\def\BB#1{\hbox{\myBB#1}}

\begin{document}

\begin{flushright}
QMUL-PH-07-14
\end{flushright}

\vspace{20pt}

\begin{center}

{\Large \bf MHV Amplitudes in $\mathcal{N} = 4$ Super Yang-Mills     }
\\
\vspace{0.3cm}
{\Large \bf  and Wilson Loops }
\vspace{12pt}
\vspace{33pt}

{\bf Andreas  Brandhuber, Paul Heslop  and Gabriele  Travaglini}%
\footnote{
{\sffamily \{\tt a.brandhuber, p.j.heslop,
g.travaglini\}@qmul.ac.uk }}

{\em Centre for Research in String Theory\\ 
Department of Physics\\
Queen Mary, University of London\\
Mile End Road, London, E1 4NS\\
United Kingdom
 }

\vspace{40pt} {\bf Abstract}

\end{center}


\noindent
It is a remarkable fact that MHV amplitudes in maximally supersymmetric Yang-Mills theory
at arbitrary loop order can be written as the product of the tree amplitude with 
the same helicity configuration and a universal, helicity-blind function of the kinematic invariants.
In this note we show how for one-loop MHV amplitudes with an arbitrary number of external legs 
this universal function can be derived using Wilson loops. 
Our result is in precise agreement with  the known expression for the infinite sequence of 
MHV amplitudes in $\cN\!=4\!$ super Yang-Mills. 
In the four-point case, we are able to reproduce  the expression of the  amplitude
to all orders in the dimensional regularisation parameter $\epsilon$.  
This prescription disentangles  cleanly  infrared divergences and finite terms, 
and leads to an intriguing one-to-one mapping between certain Wilson loop diagrams 
and (finite) two-mass easy box functions. 
\vspace{0.5cm}

\setcounter{page}{0}
\thispagestyle{empty}
\newpage


\section{Introduction}
\setcounter{footnote}{0}

Even after many years, $\cN\!=\!4$ supersymmetric Yang-Mills (SYM) remains a fascinating
theory that constantly reveals new hidden structures and symmetries. In the last few years,
substantial progress has been made in two seemingly unrelated areas of research of the
theory.

On the one hand there is the conjecture of Bern, Dixon and Smirnov (BDS) of an exponential formula for
planar  $n$-point amplitudes in $\cN\!=\!4$ super Yang-Mills (SYM) at large $N$  \cite{bds}.
According to this conjecture,  higher-loop amplitudes are determined in terms of the one-loop
amplitude together with four constants,  which depend on  the 't Hooft coupling only. 
The finite parts of these amplitudes obey a similar exponential formula: 
the all-loop finite parts are determined purely by the one-loop finite
part and  two  coupling-dependent constants. Furthermore,  one of these
constants is a well-known physical quantity, known as the cusp anomalous
dimension. 

In parallel developments,  integrability has been used,  
together with a number of further assumptions, to study the
spectrum of gauge invariant operators in $\cN\!=\!4$  SYM. 
Among the impressive outcomes of this is the conjectured 
all-orders formula of \cite{Beisert:2006ez} 
for the very same cusp anomalous dimension
appearing in the exponentiation formula,  
thus giving a tantalising  potential  link between integrability and 
amplitudes.  Calculations of four-point
amplitudes using unitarity methods  at up to a highly impressive
four-loop order have been performed~\cite{bds,4l}, thus determining the
cusp anomalous dimension   to this order (at least numerically) and 
leading to an expression for the amplitude in terms of integral functions.  
This four-loop result was then re-derived in \cite{Cachazo:2006az}, which 
further confirmed the conjecture of  \cite{Beisert:2006ez}.
Using the  assumption of pseudo-conformality, originally
observed at three \cite{Drummond:2006rz} 
and four loops \cite{4l},  the
expressions of the amplitude in terms of the 
integrals occurring at five loops has been determined~\cite{5l}.

Very recently, Alday and Maldacena have been able to apply for the first time 
the AdS/CFT correspondence to the calculation of amplitudes, 
and verified the form of  the exponentiation
of the four-point amplitude at strong coupling~\cite{am}. 
Remarkably, it turns out that the computation of
amplitudes at strong coupling is dual to the computation of the area
of a string ending on a lightlike polygonal loop embedded in the
boundary of AdS space. This,  in turn,  is equivalent to the
method for computing a lightlike polygonal Wilson loop at strong coupling
using AdS/CFT. The edges of the polygon are determined by the
external momenta of the amplitude.

In~\cite{dks} the same Wilson loop (with four lightlike segments) was considered
in weakly-coupled gauge theory
and it was shown at one loop that it reproduces the known one-loop four-point amplitude.%
\footnote{More precisely, the calculations of \cite{am} and \cite{dks}, as well as ours,  
are  insensitive to the polarisations of the particles participating 
in the scattering. In particular,  the tree-level Parke-Taylor amplitude, which 
appears as a common prefactor in the $\cN\!=4\!$ MHV amplitudes at any loop order, 
is not generated by the calculation. }
The infrared divergent pieces come from propagators stretching between adjacent
edges, and the finite part of the amplitude comes from propagators
stretching between  opposite edges.

In this paper we consider Wilson loops around arbitrary lightlike
polygons with $n$ sides at one loop,  and find precise agreement with
one-loop $n$-point MHV amplitudes in $\cN\!=\!4$ SYM (divided by the tree amplitude). 
As for the four-point case,  the infrared
part is correctly reproduced by considering propagators between
adjacent segments and the finite piece is obtained from propagators
between non-adjacent segments. This finite part is given by the sum
of the (finite) two mass easy box functions  in four dimensions with coefficients 
unity. 
This ability to separate the divergent and
finite parts is a particularly nice feature,  and enables us to
calculate finite parts purely in four dimensions without the
need of a regulator. 
On the other hand, if we use dimensional regularisation on the finite
part as well,  we reproduce the all orders in $\epsilon$ expression for 
the two-mass easy box function, and hence the complete all-orders in 
$\eps$ expression for the four-point MHV amplitude. 

The paper  is organised as follows. In section 2 we review the form of $n$-point MHV 
amplitudes  in terms of two-mass easy box functions, and discuss the BDS conjecture and the 
strong-coupling calculation of amplitudes as lightlike Wilson loops. 
In section 3 we present the one-loop calculation
of the infinite sequence of MHV amplitudes in $\cN\!=\!4$ SYM using
lightlike Wilson loops. We present our conclusions and comment on the result 
of our calculations  in section  4.

\section{One-loop MHV amplitudes in $\mathcal{N}=4$ SYM}

In this section we briefly review the expression of the infinite sequence of 
MHV amplitudes in $\cN\!=\!4$ SYM  amplitudes at one loop. 
These amplitudes were determined for the first time  in \cite{Bern:zx} using unitarity and 
collinear limits. Recently, their expression was  re-derived in \cite{bst} 
using one-loop MHV diagrams. 
In the following discussion we  suppress constant factors connected 
with dimensional regularisation when they are
not essential. 

\begin{figure}[ht]
\begin{center}
\scalebox{0.65}{\includegraphics{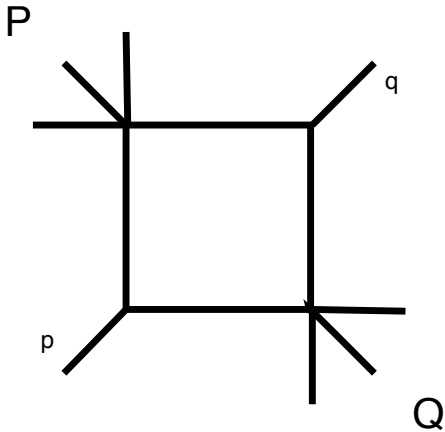}}
\end{center}
\caption{\it
A two-mass easy box function. The momenta $p$ and $q$ are null, 
whereas, in general, $P^2\neq 0$ and $Q^2\neq 0$. The cases when 
either $P^2$ or $Q^2$, or both, are also null, correspond to the 
one-mass and zero-mass boxes, obtained as smooth limits from the expression 
\eqref{2mebst} of the two-mass box function. 
}
\label{figure1}
\end{figure}

The form of the $n$-point MHV amplitudes at $L$  loops in
$\cN\!=\!4$    SYM  is remarkably simple.  
It turns out that the amplitude is  given by the tree-level amplitude, times a scalar function, 
\beq\label{fullampl}
\cA_{n}^{(L)} \ =  \cA^{\rm tree}_n\, \cM_n^{(L)}.
\eeq
At  one-loop order,  the function $\cM_n^{(1)}$  
is  expressed as a sum  of so-called two-mass easy box functions $F^{\rm 2m\,e}$ 
\cite{Bern:1993kr}, all with coefficient equal to one:%
\footnote{In \eqref{Mfunction} we are suppressing a factor of 
$
c_\G :=  {\G (1 + \e) \G^2 ( 1 -  \e)  / [  (4\pi)^{2- \e}
\G(1 - 2 \e)] }.$
}
\beq
\label{Mfunction}
 \cM_{n}^{(1)}   \ = \  \sum_{p, q}
F^{\rm 2m\,e} (p, q, P, Q)
\ .
\eeq
The main characteristic of this function, depicted in Figure \ref{figure1},  is that two opposite legs,  
$p$ and $q$, are massless, whereas the two remaining legs 
$P$ and $Q$ are massive. 
The summation in  \eqref{Mfunction} 
is such that each different two-mass easy function  appears exactly  once.%
\footnote{A more explicit way to write $ \cM_{n}^{(1)}$ is 
$ \cM_{n}^{(1)}   \!= \!\sum_{i=1}^{n} \sum_{r=1}^{[{n\over 2}]-1}
\Bigl(
1 - (1/2) \delta_{n/2  - 1, r}
\Bigr)\,
F_{n:r;i}^{\rm 2m\,e}$, where the relation to the functions introduced in \eqref{Mfunction}
and depicted in Figure \ref{figure1} is obtained by setting 
$p=p_{i-1}$, $q=p_{i+r}$, and $P=p_i + \cdots +
p_{i+r-1}$. }

A compact   form of the two-mass easy box function containing only four dilogarithms was first 
derived in \cite{Duplancic:2000sk}. This form  was found independently in \cite{bst} in the context of 
MHV diagrams, where an analytic proof of its equivalence  with the conventional expression of  
e.g.~\cite{Bern:1993kr}  was given. 
Expressing the two-mass easy box  as a function of the kinematic invariants 
$s := (P+p)^2$,  $t := (P+q)^2 $ and $P^2$,  $Q^2$, with $p+q+P+Q=0$, 
it reads 
 \beqa
\label{2mebst}
\hspace{-0.5cm}
 F^{\rm 2me} (s,t,P^2, Q^2) &=& 
-\frac{1 }{\e^2}\Big[ \left({-s\over \mu^2}\right)^{-\e} \, + \, \left({-t\over \mu^2}\right)^{-\e} \, - \,
  \left({-P^2\over \mu^2}\right)^{-\e}\, - \,
\left({-Q^2\over \mu^2}\right)^{-\e}\Big]  \\ \cr
\nonumber
   &+& \Li(1-aP^2)\, + \, \Li(1-aQ^2)  \, -\,  \Li(1-as)
\,  -\,  \Li(1-at)
\ , 
\eeqa
where
\beq 
\label{adef} 
a \ = \
\frac{P^2+Q^2-s-t}{P^2Q^2-st} \ = \
\frac{u}{P^2Q^2-st}
\ . 
\eeq

For later use, we also quote the all-orders in $\epsilon$ expression of this function
\cite{ftt}, 
\beqa
\label{puzzola}
&& \hspace{-0.7cm}
F^{\rm 2me} (s, t, P^2, Q^2) =
-{1\over \e^2}
\left[
 \Big( {-s \over \mu^2} \Big)^{-\e} \,
\, + \,
\Big( {-t \over \mu^2} \Big)^{-\e}
\, - \, 
\Big( {-P^2 \over \mu^2} \Big)^{-\e}
\, - \, 
 \Big( {-Q^2 \over \mu^2} \Big)^{-\e} \,
\right. 
\\ 
&& \hspace{-0.3cm} \left. + \, 
\Big( {a \m^{2} \over 1-aP^2} \Big)^{\e} \,
\mbox{}_{2}F_1 \left( \e, \e, 1+ \e, {1 \over 1 - a P^2 }\right)
\, + \,
\Big( {a \m^{2} \over 1-aQ^2} \Big)^{\e} \,
\mbox{}_{2}F_1 \left( \e, \e, 1+ \e, {1 \over 1 - a Q^2 }\right)
\right.
\nonumber 
\\ [6pt] \cr
&& \hspace{-0.3cm}  - \,
\left.
\Big( {a \m^{2} \over 1-as} \Big)^{\e} \,
\mbox{}_{2}F_1 \left( \e, \e, 1+ \e, {1 \over 1 - a s }\right)
\, - \,
 \Big( {a \m^{2} \over 1-at} \Big)^{\e} \,
\mbox{}_{2}F_1 \left( \e, \e, 1+ \e, {1 \over 1 - a t }\right)
\right]
\ . 
\nonumber
\eeqa

One important property of the MHV amplitudes  is that they do  not have 
multiparticle singularities. In particular, 
we note that, although each box function \eqref{2mebst} contains
poles in $\eps$ associated to multiparticle invariants, after performing the sum 
\eqref{Mfunction} the infrared divergent terms only involve two-particle invariants, 
\beq
\label{MIR}
 \cM_{n}^{(1)}\left|_{\rm IR}\right.    \  = \ 
- {1\over \eps^2} \sum_{i=1}^n  \left( {-s_{i i+1}\over  \mu^2}\right)^{- \eps} 
\ , 
\eeq
with $s_{i j} := (p_i + p_{i+1})^2$. 

Recently, Bern, Dixon and Smirnov (BDS)  proposed a remarkably 
simple conjecture for  the resummation at all loops of the planar MHV amplitudes 
in $\cN\!=\!4$   SYM calculated at weak coupling \cite{bds}. 
This conjecture,  based on explicit calculations at two loops 
\cite{abdk}, was verified in \cite{bds} up to three loops in the four-point case,  
and in \cite{2l5pt} up to two loops at five points.  
Explicit expressions of the four-point amplitudes at four and five loops were 
recently presented in \cite{4l} and \cite{5l}, respectively, and will allow for precise tests 
of the conjecture at four and five loops once the relevant integral functions have been evaluated 
to the necessary degree of accuracy in $\epsilon$. 

The form of the BDS conjecture is inspired by the soft and collinear behaviour 
of amplitudes in gauge theory \cite{ir1,ir2,ir3,ir4,ir5,ir6,ir7,ir8}, 
and  is expressed  by  
\beq
\label{bds}
\cM_n \ := \ 1 + \sum_{L=1}^{\infty} a^L \cM_{n}^{(L)} (\epsilon )  \ =  \ 
\exp \Big[ \sum_{L=1}^{\infty} a^L  \Big( f^{(L)} (\epsilon) \cM_{n}^{(1)}  + C^{(L)} + E_n^{(L)}(\epsilon )\Big) 
\Big] 
\ , 
\eeq
where $a=[{g^2 N/ (8 \pi^2)}] (4\pi e^{-\gamma})^\eps$. Here $f^{(L)}(\epsilon ) = f_0^{(L)} + f_1^{(L)} \epsilon + f_2^{(L)} \epsilon^2$ is a set of functions, 
one at each loop order, which make their appearance in the exponentiated all-loop expression 
for the  infrared divergences in generic amplitudes in dimensional regularisation \cite{ir6}. 
In particular, $f_0^{(L)} = \gamma_{K}^{(L)} / 4$, where $\gamma_{K}$ is the cusp anomalous dimension
(related to the anomalous dimension of twist-two operators of large spin). 
An important point of the conjecture is that the constants $C^{(L)}$ do not depend on 
kinematics or on the number of particles $n$. The non-iterating contributions 
$E_n^{(L)}$ vanish as $\epsilon \to 0$ and depend explicitly on $n$. 

BDS also propose a very interesting form for the finite remainders of the MHV amplitude, 
given by 
\beq \label{finite}
\mathcal{F}_n \ =  \exp \Big[ {1\over 4} \gamma_K \, F^{(1)}_n (0)  + C \Big] 
\ , 
\eeq
where $F^{(1)}_n (0)$ is the finite remainder  of $ \cM_{n}^{(1)}$ as defined in \cite{bds}.

Motivated by the BDS conjecture,  Alday and Maldacena have managed to
reproduce the exponential 
formula \eqref{bds} at strong  coupling for the four-point case 
using the AdS/CFT correspondence~\cite{am}. The AdS dual description of a
planar colour-ordered amplitude in $\cN\!=\!4$  SYM is given by a classical
open string worldsheet ending 
on a brane placed in the far infrared region of AdS space. Specifically, 
using coordinates in which the metric of AdS
space is
\begin{equation}
 ds^2={R^2 }\left(dx^2+dz^2 \over z^2 \right)
 \ , 
\end{equation}
then  the boundary of AdS space is at $z=0$ and the infrared brane
sits at $z=\infty$.

It is convenient however to use  a
T-dual description of this string configuration. This T-duality maps
the AdS space into a new space with coordinates $(y^{\mu}, r)$ which
has the metric
\begin{equation}
d\tilde s^2=R^2  \left(dy^2+dr^2 \over r^2\right)
\ , 
\end{equation}
where $r={R^2/  z}$. We see that the new space is again AdS,  but now the infrared region
and the boundary have been inverted. 
Therefore,  the brane is located on the boundary in the new coordinates.
Furthermore,  the momenta of the particles $p_i$ are expressed as differences of 
dual,  or region momenta $y_i$ \cite{thooft},    $(2 \pi ) p_i \ = \ y_i - y_{i+1}$. 
The calculation of the amplitude thus becomes that of finding the classical action 
 $S_{\rm cl}$  of a 
string worldsheet  whose boundary is a polygon
with vertices 
$y_i$ lying within the AdS boundary,
\begin{equation}
\cM_n\sim   e^{ i S_{\rm cl}}.
\end{equation}
For the four-point amplitude,  the corresponding string solution can be
determined \cite{am} giving  $iS_{\rm cl}={\rm{div}} +({\sqrt{\lambda} / 
  8\pi}) \,\log^2\left(s /  t \right)+C$ where $\rm{div}$ represents
divergent terms. This agrees precisely with the BDS conjecture~(\ref{finite}) and
predicts $\gamma_K={\sqrt{\lambda} / 4\pi}$ at large $\lambda=g^2 N$,  
in agreement with string calculations~\cite{Gubser:2002tv,Frolov:2002av}.

Now,  the minimal area of a string ending on a path in the
boundary of AdS space gives the vacuum expectation value of the 
Wilson loop over the same path in the CFT at strong coupling \cite{rey,maldawil}. 
A subtlety arising in the case at hand is the presence of singular points or cusps in the path, 
which lead to divergences \cite{am,Buchbinder:2007hm}. Nevertheless the
 divergences can be regularised by dimensional reduction even in the
 string calculation.
Therefore, at least at strong coupling there is evidence for  a dual description
of amplitudes as Wilson loops. 
In the next section we will consider this possibility at small coupling.  
An important point to note here is that the  string calculation
does not depend on the species or helicities of the particles in the
amplitude. These are subleading terms which would require $\alpha'$
corrections \cite{am,valya}. 
Our Wilson loop calculation is also insensitive to the helicities
of the scattered particles; thus, similarly to the  Alday-Maldacena result,  
it does not generate the tree-level Parke-Taylor amplitude. 

One mysterious and intriguing consequence of this dual description of amplitudes as
Wilson loops is the  unexpected appearance of  conformal symmetry.  
Wilson loops of smooth paths in $\cN\!=\!4$  SYM are conformally invariant objects 
(modulo an anomaly  which does not depend either on the shape or size of the loop \cite{esz,dg}). 
However here the Wilson loop is divergent, since the path is not
smooth, and regularisation spoils the conformal symmetry. 
Nevertheless, a similar pseudo-conformality seems to appear at weak coupling
where all the integrals contributing to four-point MHV diagrams can be
determined by rewriting them using the region momenta and appealing to
off-shell conformality~\cite{Drummond:2006rz,4l,dks}. Furthermore at
four points all
conformal integrals of a certain type and with certain singular
properties appear with coefficients $\pm 1$. 
The Wilson loop picture would seem to suggest that this pseudo-conformal invariance
should continue for $n$-point functions. This point clearly deserves further 
investigation.  


\section{MHV amplitudes from a Wilson loop calculation}
In this section we calculate one-loop corrections to the vacuum expectation value 
of a particular Wilson loop. In 
$\cN\!=4\!$ SYM, the appropriate operator takes  the form (suppressing fermions) 
\cite{wil1,wil2,wil3}
\beq
\label{wil}
W[ \cC]  \ := \ {\rm Tr} \, \cP \exp \left[ i g\oint_{\cC} \! d\tau \Big(  A_\mu (x(\tau )) \dot{x}^{\mu} (\tau ) + 
\phi_i  (x (\tau)) \dot{y}^i (\tau) \Big) \right] 
\ , 
\eeq
where the $\phi_i$'s are the six scalar fields of $\cN\!=\!4$ SYM, and 
$( x^\mu (\tau), y^i (\tau))$ parametrise the loop $\cC$. 
Importantly, when $\dot{x}^2  = \dot{y}^2$,  the Wilson loop  \eqref{wil} is locally
supersymmetric. 
The specific form of the contour $\cC$ we choose is  dictated by the gluon momenta 
$p_1, \cdots , p_n$. 
Specifically, the segment associated 
to momentum $p_i$ will be delimited by  $k_i$ and  $k_{i+1}$, 
\beq
p_i \ := \ k_i - k_{i+1}
\ ,  
\eeq
and  will be parametrised as 
$k_i (\tau_i ) := k_i + \tau_i (k_{i+1} - k_i)= k_i - \tau_i p_i$, 
$\tau_i \in [0,1]$. 
Momentum conservation  $\sum_{i=1}^{n} p_i = 0$ implies that the contour is closed. 
In addition we set $\dot{y}^i = 0$ which makes the Wilson loop locally supersymmetric, 
as the gluon momenta, and hence the segments of the contour are null.%
\footnote{We thank George Georgiou for discussions on this point.}
However, we notice that each segment of the loop preserves a different subset 
of supersymmetries, therefore supersymmetry is broken globally.  

We notice that the coordinates $k_i$ can be interpreted as dual, or  region momenta
\cite{thooft}.  Indeed, for any planar diagrams one can  express the momentum carried by a 
line as the difference of  the momenta of the two regions of the plane separated by 
the segment. These coordinates have been used in the context of amplitude calculations by Thorn and collaborators in \cite{thorn1,thorn2} and, more recently, in \cite{bstz} in the context of MHV diagrams. 
They are also the T-dual coordinates introduced in \cite{am} which
determine the classical string solutions, as discussed earlier, 
and appear in the conformal integrals discussed in  \cite{Drummond:2006rz}.

The four-particle case was recently addressed in \cite{dks}, where it was found that the result of a one-loop 
Wilson loop calculation reproduces the four-point MHV amplitude in $\cN\!=\!4$ SYM. 
Here we extend this result 
in two directions. First, we derive the four-point MHV amplitude to all-orders in the dimensional regularisation 
parameter $\epsilon$. Secondly, we show that this striking agreement persists for an MHV amplitude with an arbitrary number of external particles.

Three different classes of diagrams give one-loop corrections to the Wilson loop.%
\footnote{Notice that, for a Wilson loop bounded by gluons, we can only exchange 
gluons at one loop.}
In the first one,  a gluon stretches between points belonging to the same segment. It is immediately seen 
\cite{dks} that these diagrams give a vanishing contribution. 
In the second class of diagrams, a gluon stretches between two adjacent segments meeting at a cusp. 
Such diagrams  are ultraviolet divergent and were calculated long ago 
\cite{polyakov,cu1,cu2,cu3,cu4,cu5,cu6,cu7}, specifically in \cite{cu6,cu7} 
for the case of  gluons attached to lightlike segments. 

\begin{figure}[ht]
\begin{center}
\scalebox{0.40}{\includegraphics{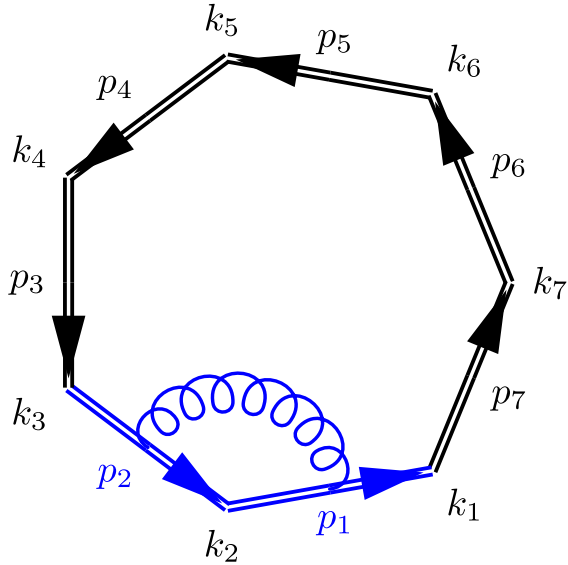}}
\end{center}
\caption{\it
A one-loop correction to the Wilson loop, where the gluon stretches between two 
lightlike momenta meeting at a cusp. Diagrams in this class provide the infrared-divergent terms 
in the $n$-point scattering amplitudes, given in \eqref{MIR}. 
}
\label{figure2}
\end{figure}

In order to compute these diagrams, we will use the gluon propagator 
in the dual configuration space, which in $D=4-2 \epsuv$ dimensions is 
\beqa
\D_{\mu \nu} (z) & := & 
- {\pi^{2 - {D\over 2}} \over 4 \pi^2}
\Gamma \Big({D\over 2} - 1 \Big)
 {\eta_{\mu \nu} \over (-z^2+ i \varepsilon)^{{D\over 2} - 1}}
\\ \nonumber 
&=&  
- {\pi^{\epsuv} \over 4 \pi^2}\Gamma (1-\epsuv)  \ {\eta_{\mu \nu} \over 
(-z^2+ i \varepsilon)^{1-\epsuv }}
\ . 
 \eeqa
A typical diagram in the second class is pictured in Figure \ref{figure2}. 
There one has $k_1 (\tau_1) - k_2 (\tau_2 ) = p_1 (1 - \tau_1 ) + p_2 \tau_2$, 
where we used
$p_1= k_1 - k_2 $ and $p_2 = k_2 - k_3$. 
The cusp diagram then gives%
\footnote{After changing variables $1-\tau_1 \to \tau_1$.}
\beqa
\label{cuspintegral}
&&-(i g \tilde \mu^{\epsuv})^2 \, {\Gamma (1 - \epsuv) \over 4 \pi^{2 - \epsuv}} \ 
\int_{0}^{1}\!\!d\tau_1 d\tau_2 \ {(p_1 p_2 ) \over [ - \big(p_1 \tau_1 + p_2 \tau_2\big)^2]^{1-\epsuv}} 
\nonumber \\ 
&  = & 
 -(i g \tilde \mu^{\epsuv})^2 \, {\Gamma (1 - \epsuv) \over 4 \pi^{2 - \epsuv}} \ 
 \left[-  {1\over 2} {(-s)^{\epsuv}\over \epsuv^2 } \right] \ . 
\eeqa
The UV divergence should be interpreted as a divergence at small differences of 
region momenta, i.e.~momenta, hence we interpret it as an infrared singularity 
in momentum space. Notice that $\epsuv >0$, in order to regulate the divergence
in \eqref{cuspintegral}. Furthermore the scale used in the Wilson loop
calculation is related to the scale used to regulate the amplitudes $\mu$ as
$ \tilde \mu=(c \mu)^{-1}$ (the precise coefficient $c$ in front of $\mu$  can be reabsorbed into an 
appropriate redefinition of the coupling constant).

\begin{figure}[ht]
\begin{center}
\scalebox{0.40}{\includegraphics{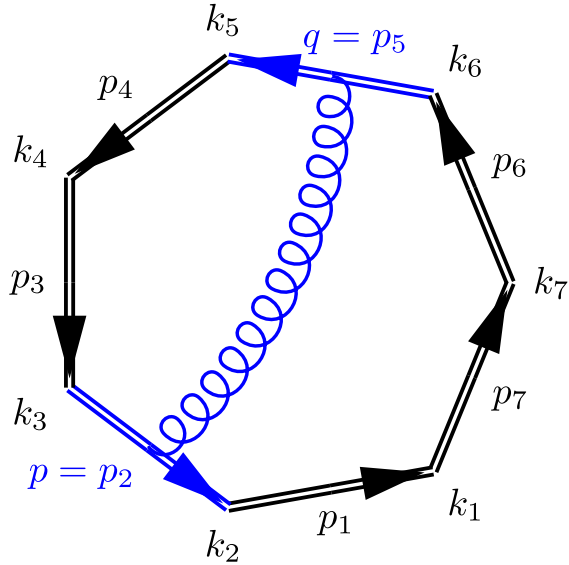}}
\end{center}
\caption{\it
Diagrams in this class -- where a gluon connects two non-adjacent segments -- are finite, and give a contribution equal to the finite part of a two-mass easy box function $F^{\rm 2me} (p, q, P, Q)$, 
second line of \eqref{2mebst}. 
$p$ and $q$ are the massless legs of the two-mass easy box, and correspond to the segments 
which are connected by the gluon. The diagram depends on the other gluon momenta only through the 
combinations $P$ and $Q$.  
}
\label{figure3}
\end{figure}

The last class of diagrams consists of diagrams where the gluon connects non-adjacent segments, such as 
that pictured in Figure \ref{figure3}. 
We denote by $p$ and $q$ the momenta carried by the two segments, and calculate the 
one-loop contribution due to the gluon exchange. We also set 
$p := k_p - k_{p+1}$, $q:= k_q - k_{q+1}$. 
The gluon propagator is a function  of 
\beq
\big( k_p (\tau_p ) - k_{q} ( \tau_{q}) \big)^2 \ = \ 
\Big(\sum_{i= p}^{q-1} (k_i - k_{i+1})  - \tau_p p + \tau_q q \Big)^2 
\ = \ \Big( p ( 1 - \tau_p ) + q \tau_q + \sum_{i = p+1}^{q-1} (k_i - k_{i+1} ) \Big)^2
\eeq
We recognise that $\sum_{i = p+1}^{q-1} (k_i - k_{i+1} ) = P$ is  the sum of the momenta between 
$p$ and $q$, where, in general, $P^2 \neq 0$. 
Hence 
\beqa
 \big( k_p (\tau_p ) - k_{q} ( \tau_{q}) \big)^2 
& = & P^2 + 2 (p P) (1 - \tau_p) + 2 (q P) \tau_q + 2 (p q ) \tau_p \tau_q 
\\ 
&= & P^2 + (s- P^2)  (1-\tau_p) + (t - P^2)   \tau_q + (-s - t + P^2 + Q^2)  \tau_p \tau_q 
\ , 
\nonumber 
\eeqa
where we have re-expressed the result in terms of 
the invariants $s := (P+p)^2$,  $t := (P+q)^2 $ defined earlier. 
We can also introduce $u:= -s - t + P^2 + Q^2$. 

The one-loop diagram in Figure \ref{figure3} is equal to 
\beq
- (i g \tilde \mu^\epsuv)^2 \,{1\over 2} \,  {\Gamma (1 - \epsuv) \over 4 \pi^{2 - \epsuv}} \ 
\cF_\epsilon  (s, t, P, Q)
\ , 
\eeq
where 
$\cF_\epsilon  (s, t, P, Q)$
is the following  two-dimensional integral,%
\footnote{In the following we set $\eps:= - \epsuv <0$, where $\eps$ will correspond to 
the usual infrared regulator.}
\beq
\label{twodim2mef}
\cF_\epsilon  (s, t, P, Q)  =  
\int_{0}^{1} \! d\tau_p d \tau_q \ 
{ P^2 + Q^2- s - t \over  
[ - \big( P^2 + (s- P^2)  \tau_p + (t - P^2)   \tau_q + (-s - t + P^2 + Q^2)  \tau_p \tau_q\big)]^{1 + \epsilon} }
\ . 
\eeq

The integral is finite in four dimensions.  We begin by calculating it 
in four dimensions setting $\epsilon = 0$ (and will come back later 
to the calculation  for  $\epsilon \neq 0$).  
In this case, the result is 
\beqa
\label{r1}
\cF_{\epsilon = 0}  (s, t, P, Q) & = &
 \Li (a s)  + \Li (a t ) -  \Li ( a P^2 ) -  \Li ( a Q^2 ) 
\\ 
  &+&
 \log s \log {(P^2 - s) (Q^2 - s) \over P^2 Q^2 - st}  +   \log t \log {(P^2 - t) (Q^2 - t) \over P^2 Q^2 - st} 
 \nonumber \\ 
 &-&
\log P^2  \log { - (P^2 - s) (P^2 - t) \over P^2 Q^2 - st} - 
 \log Q^2  \log { - (Q^2 - s) (Q^2 - t) \over P^2 Q^2 - st}
\ , 
\nonumber
\eeqa
where $a$ is defined in \eqref{adef}. 
Using Euler's identity 
\beq
\Li (z) = - \Li (1-z) - \log z \log (1-z) + {\pi^2 \over 6 }
\ , 
\eeq
and noticing that  \cite{bst} 
$(1-as)(1-at) / [ (1-aP^2)(1-aQ^2)] = 1$, 
we can rewrite 
\beqa
&&  \Li (a s)  + \Li (a t ) -  \Li ( a P^2 ) -  \Li ( a Q^2 )  \ = \  
\\ 
&-&    \Li (1-a s)  -  \Li (1- a t ) +  \Li (1-  a P^2 ) +   \Li ( 1- a Q^2 )  
\nonumber \\ 
&- & \log s \log (1-a s) -  \log t \log (1-a t) + 
 \log P^2 \log (1-a P^2) +  \log Q^2 \log (1-a Q^2)
\ . 
\nonumber 
\eeqa
Upon making use of the relations \cite{bst}
\beqa
1 - a s &=& {(s - P^2) (s - Q^2) \over P^2 Q^2 - st} 
 \ , \qquad \ \,
 1 - a t  \ = \    {(t - P^2) (t - Q^2) \over P^2 Q^2 - st}\ , 
\cr
1 - a P^2  & = &  - {(s - P^2) (t  - P^2) \over P^2 Q^2 - st}
\ , 
\quad 
1 - a Q^2  \  = \    - {(s - Q^2) (t  - Q^2) \over P^2 Q^2 - st}
\ , 
\eeqa
we see that the terms in \eqref{r1} containing logarithms cancel,  
and we are left with  
\beq
\label{conclu}
\mathcal{F}_{\epsilon = 0} = 
 -    \Li (1-a s)  -  \Li (1- a t ) +  \Li (1-  a P^2 ) +   \Li ( 1- a Q^2 )  
\ . 
\eeq
We conclude that  $\mathcal{F}_{\epsilon = 0}$ is precisely equal to the finite part of the 
two-mass easy box function --  second line of  \eqref{2mebst}. 
Finally, summing over all possible pairs of non-adjacent segments reproduces precisely 
the sum over box functions  in \eqref{Mfunction}. 

In the four-point case, $a\left|_{P^2 = Q^2 = 0}\right. = 1/s + 1/t$, and using 
\beq
\Li (z) + \Li  \left({1 \over z}\right)  + {1\over 2} \log^2 ( - z) + {\pi^2 \over 6} = 0
\ , 
\eeq
one immediately finds \cite{dks}
\beq  
\mathcal{F}_{\epsilon = 0}\left|_{P^2 = Q^2 = 0}\right. \ = \ 
{1\over 2} \log^2 \left( {s \over t} \right) + {\pi^2 \over 2}
\ , 
\eeq
in complete agreement with the finite parts of the zero-mass box function
(in the normalisations of \cite{Bern:zx}).

Finally, we discuss the calculation at  $ \eps \neq 0$. In this case one finds that 
\beqa
\label{333}
&&\mathcal{F}_{\epsilon} \, = \, 
-{ 1 \over \e^2}
\\ [6pt]\nonumber
&&
\hspace{-0.4cm} 
\cdot
\left[
\Big( {a \over 1-aP^2} \Big)^{\e} \,
\mbox{}_{2}F_1 \left( \e, \e, 1+ \e, {1 \over 1 - a P^2 }\right)
\, + \,
\Big( {a  \over 1-aQ^2} \Big)^{\e} \,
\mbox{}_{2}F_1 \left( \e, \e, 1+ \e, {1 \over 1 - a Q^2 }\right)
\right.
\\ [6pt] \cr
&& \hspace{-0.4cm}  - \,
\left.
\Big( {a  \over 1-as} \Big)^{\e} \,
\mbox{}_{2}F_1 \left( \e, \e, 1+ \e, {1 \over 1 - a s }\right)
\, - \,
 \Big( {a  \over 1-at} \Big)^{\e} \,
\mbox{}_{2}F_1 \left( \e, \e, 1+ \e, {1 \over 1 - a t }\right)
\right]
\ . 
\nonumber
\eeqa
This result is in precise agreement with the finite part of the all-orders in $\epsilon$ two-mass easy box function, second and third line  of  \eqref{puzzola}. 
Notice that the expression in \eqref{333} is finite as $\eps\to 0$. 

For $n>4$, simply replacing the $\eps=0$ expression of the box functions with their all-orders in $\eps$ 
expression does not provide us with a complete, all-orders in $\eps$ expression for the amplitude, as 
there are additional  $n$-gon integrals which vanish as  $\eps\to 0$,  which are not included. 
The four-point case is an exception, and in this case our calculation reproduces the expected 
all-orders in $\eps$ result. Combining the infrared-divergent and finite terms, 
our result is 
\beq
\label{4ptymeps}
\hspace{-0.2cm}\cM^{(1)}_4 (\eps)   =   - \frac{2}{\epsilon^2} \left[
\left({-s\over \mu^2} \right)^{- \epsilon}
{}_2F_{1}\left( 1, - \epsilon, 1- \epsilon, 1+\frac{s}{t} \right) +
\left({-t\over \mu^2}\right)^{- \epsilon}
{}_2F_{1} \left( 1, - \epsilon, 1- \epsilon, 1+\frac{t}{s} \right) \right] \, , 
\eeq
in agreement with \cite{gsb}. 

Finally, we would like to mention that it is easy to see that splitting amplitudes and 
soft functions can also be derived at one loop using Wilson loops. 
Furthermore, they have an interesting  interpretation in terms of 
the geometry of the contour; for instance, splitting amplitudes 
arise from two adjacent segments becoming nearly parallel 
to each other, therefore merging into a single segment.

\section{Conclusions}

We found that a strikingly simple  one-loop gluon exchange calculation 
of a  Wilson loop, whose (closed) boundary is
defined by a set of lightlike segments $p_1, \ldots , p_n$,  reproduces  the 
$n$-point one-loop  MHV amplitudes in $\cN\!=\!4$  SYM 
(divided by the tree-level amplitude).
We would like to comment on this surprising result.

{\bf 1.} One of the important features of the calculation presented in this paper 
is that it neatly separates  the infrared-divergent terms from the finite parts. 
The latter can then be derived working directly in four dimensions, which turns out  to be 
a key calculational advantage. For this reason, we are hopeful that  this procedure 
could allow for a direct check of the exponentiation of the finite remainders   \cite{bds}. 
The perspective of deriving  a field-theoretical proof of the all-loop expressions of \cite{bds} 
from Wilson loops  -- possibly using  the non-abelian exponentiation theorem \cite{gatheral,taylor}  -- 
is an exciting one.

{\bf 2. }
In \cite{bst},  it was speculated that the two-mass easy box functions should emerge 
naturally  as Feynman diagrams of the perturbative description of a (possibly string) theory. 
This was motivated by the observation that the MHV amplitudes in 
$\cN\!=\!4$  SYM  at one loop 
are written as sum of box functions, each appearing with coefficient one.  
In this paper, we have found  that the Wilson loop calculation gives a precise, 
one-to-one mapping of Wilson loop diagrams 
to the {\it finite part} of  two-mass easy box functions  
(or, in specific cases,  the degenerate  one-mass and zero-mass functions).  
The massless legs of the box function, $p$ and $q$, are simply those to which the gluon is attached 
(see Figure \ref{figure3}). The calculation is only sensitive to $p$, $q$, and the {\it sum } $P$ 
of the momenta between $p$ and $q$. 
Therefore, the Wilson loop calculation seems to have provided such a description 
where a two-mass easy box is a specific Wilson loop diagram 
(notice that for the diagram with a gluon connecting segments $p$ and  $q$, 
the remaining part of the loop could be deformed to the shape of a (generically) two-mass 
box function).%
\footnote{We also note in passing that we have found a representation of the finite part of 
two-mass easy box functions in terms of a very simple two-dimensional integral, 
see \eqref{twodim2mef} and \eqref{conclu}.}
It is tempting to speculate that the observation of \cite{4l} that  even at higher loops 
the MHV amplitudes  are expressed as sums of (conformal) integrals, 
each appearing with coefficients $\pm1$, could be explained in terms of higher-loop 
Wilson loop diagrams. 
 
{\bf 3. }
We believe  that the agreement found in this paper 
between our Wilson loop calculation  and the $\cN\!=\!4$  MHV amplitude with an arbitrary number 
of points lends support  to the conjecture that the appropriate strongly-coupled string theory 
calculation  at $n$ points will confirm the BDS conjecture for the full exponentiated expression of the 
$n$-point MHV amplitude.


\newpage

\section*{Acknowledgements}

It is a pleasure to thank James Drummond, George Georgiou, Valeria Gili, 
Gregory Korchemsky, Sanjaye Ramgoolam, Rodolfo Russo, 
Emery Sokatchev, Bill Spence  and Costas Zoubos for discussions. 
We would like to thank PPARC for support under a
Rolling Grant PP/D507323/1 and the Special Programme Grant PP/C50426X/1.
The work of PH is supported by an EPSRC Standard Research Grant EP/C544250/1.  
GT is supported by an EPSRC Advanced Fellowship EP/C544242/1
and by an EPSRC Standard Research Grant EP/C544250/1.


\end{document}